\documentstyle[aaspp4,flushrt]{article}
\def\ro{r_{\rm out}}
\begin{document}

\title{Spectral Models of Convection-Dominated Accretion Flows}
\author{Gregory H. Ball, Ramesh Narayan} \affil{Harvard-Smithsonian
Center for Astrophysics} \affil{60 Garden Street, Cambridge MA 02138,
U.S.A} \author{and} \author{Eliot Quataert} \affil{Institute for
Advanced Study, Einstein Drive, Princeton, NJ 08540}

\begin{abstract}
For small values of the dimensionless viscosity parameter, namely
$\alpha\lesssim 0.1$, the dynamics of non-radiating accretion flows is
dominated by convection; convection strongly suppresses the accretion
of matter onto the central object and transports a luminosity $\sim
10^{-3}-10^{-2} \dot M c^2$ from small to large radii in the flow.  A
fraction of this convective luminosity is likely to be radiated at
large radii via thermal bremsstrahlung emission.  We show that this
leads to a correlation between the frequency of maximal bremsstrahlung
emission and the luminosity of the source, $\nu_{\rm peak} \propto
L^{2/3}$.  Accreting black holes with X-ray luminosities $10^{-4}
L_{Edd}\gtrsim L_X(0.5-10{\rm keV}) \gtrsim 10^{-7}L_{Edd}$ are
expected to have hard X-ray spectra, with photon indices
$\Gamma\sim2$, and sources with $L_X\lesssim 10^{-9}L_{Edd}$ are expected to
have soft spectra, with $\Gamma\sim3.5$.  This is testable with {\it
Chandra} and {\it XMM}.

\end{abstract}

\section{Introduction}
\label{sec:Introduction}

At luminosities less than a few per cent of the Eddington luminosity,
black holes can accrete via an advection-dominated accretion flow, or
ADAF (Ichimaru 1977; Rees et al. 1982; Narayan \& Yi 1994, 1995;
Abramowicz et al. 1995; see Kato, Fukue \& Mineshige 1998 and Narayan,
Mahadevan \& Quataert 1999 for reviews).  Analytical calculations have
shown that such flows should be convectively unstable (Narayan \& Yi
1994, 1995; see also Begelman \& Meier 1982); the instability has been
confirmed in numerical simulations (Igumenshchev, Chen \& Abramowicz
1996; Igumenshchev \& Abramowicz 1999, 2000; Stone, Pringle \&
Begelman 1999; Igumenshchev, Abramowicz \& Narayan 2000).

Igumenshchev \& Abramowicz (1999, 2000) found that convection is
strong whenever the viscosity parameter $\alpha$ is small, roughly
$\alpha \lesssim 0.1$.  Such ``convection-dominated accretion flows,''
or CDAFs, have a very different structure than ADAFs (Stone et
al. 1999).  The density of the accreting gas varies as $\rho\propto
R^{-1/2}$ (where $R$ is the radius) rather than $\rho \propto
R^{-3/2}$, and the mean radial velocity varies as $v\propto R^{-3/2}$
rather than $v\propto R^{-1/2}$ (Narayan, Igumenshchev \& Abramowicz
2000, hereafter NIA).

NIA showed that, in the numerical simulations, the Reynolds stress due
to convection is negative, which implies that convection moves angular
momentum {\it inwards} rather than outwards.  This property of
convection was discussed in the context of thin accretion disks by Ryu
\& Goodman (1992) and Stone \& Balbus (1996).  In the case of a CDAF,
convection is so strong that the angular momentum it transports
inwards nearly cancels the normal outward transport by viscosity (NIA;
Quataert \& Gruzinov 2000).  As a result, one has a nearly static
accretion flow in which most of the gas circulates in convective
eddies rather than accreting onto the central object.  For fixed
boundary conditions at large radii the mass accretion rate in a CDAF
is thus much smaller than in a non-convecting ADAF or a Bondi flow.

In this paper we highlight observational consequences of CDAF models,
focusing in particular on spectra.  Some of our results are similar to
those of Di Matteo et al. (1999, 2000; DM) and Quataert \&
Narayan (1999a; QN) who followed up the proposal of Blandford \&
Begelman (1999; see also Narayan \& Yi 1994, 1995) that a significant
fraction of the mass in an ADAF would be lost to an outflow/wind,
rather than accreting onto the central object.  The importance of
outflows can be parametrized by a radial density profile, $\rho
\propto R^{-3/2 + p}$, with $p \ \epsilon \ [0,1]$.  The density
profile in a CDAF is equivalent to $p = 1$.  Thus DM's and QN's
spectral models, which considered various values of $p$, capture many
of the features of the CDAF spectra calculated here; this is discussed
in more detail in \S4.

Convection in CDAFs transports a luminosity $L_c \sim 10^{-3}-10^{-2}
\dot M c^2$ from small to large radii; the energy is supplied by the
small amount of mass accreting onto the black hole.  What happens to
this energy at large radii?  NIA and Igumenshchev \& Abramowicz (2000)
suggested that some, perhaps most, of the energy might be radiated
from the outer regions of the CDAF as thermal bremsstrahlung emission.
We present model spectra for such flows.

In the next section we present analytical expressions for the
luminosity and bremsstrahlung spectrum of CDAFs (\S2).  We then show
more detailed numerical calculations (\S3.1) and compare CDAF models
with ADAF models (\S3.2).  In \S4 we summarize and emphasize some
implications of our analysis.

\section{Self-Similar Scalings for Bremsstrahlung Luminosity and Spectrum}
\label{sec:Self-Similar Scalings}

In this section we use the self-similar CDAF solution described by
Narayan et al. (2000) and Quataert \& Gruzinov (2000).  We employ
Schwarzschild units for the radius, i.e. $r = R/R_{\rm S}$, where
$R_{\rm S} = 2GM/c^{2}= 2.95\times10^5m$ cm and $m$ is the black hole
mass in solar units.  We assume that the accretion flow extends from
an outer radius $r_{\rm out}$ down to an inner radius $r_{\rm in} = 1$
(to model a Schwarzschild black hole).  The sound speed of a CDAF is
nearly virial, $c^2_s \approx 0.37 c^2/r$, and the temperature is $T
\equiv T_0/r \approx 10^{12}{\rm K}/r$; the vertical scale height is
thus of order the radius, $H \approx 0.6 R$.

The density of gas in a CDAF is given by
\begin{equation}\label{eq:rho scaling}
\rho =\rho _{0}r^{-1/2}=\rho_{\rm out}\left({r\over
r_{out}}\right)^{-1/2}, \qquad \rho _{0}=\rho_{\rm out}r^{1/2}_{\rm
out}.
\end{equation}
We assume that the CDAF is in steady state with a constant mass
accretion rate $\dot M$. Since $\rho$ varies as $r^{-1/2}$, the radial
velocity must scale as $v \propto r^{-3/2}$.  We normalize the
velocity such that it is equal to $c$ at $r_{\rm in}$.

As noted in \S1, convection in CDAFs transports energy from small to
large radii.  We write the convective luminosity as $L_c \equiv
\epsilon_c \dot M c^2$, where $\epsilon_c \sim 10^{-2}-10^{-3}$ is the
convective ``efficiency."  Let a fraction $\eta_c$ of this energy
be radiated at large radii by the gas in the CDAF; this radiation
comes out as thermal bremsstrahlung emission.  The bremsstrahlung
luminosity of the CDAF, $L_{\rm CDAF}$, then satisfies
\begin{equation}
L_{\rm CDAF}=\eta_cL_{\rm c}=\eta_c
\epsilon _{c}\dot{M} c^{2}.\label{equality}
\end{equation}
For a given $r_{\rm out}$, this relation fixes the value of $\rho_{\rm
out}$ and thus $\dot M$ (see below).  Equivalently, for a given $\dot
M$, equation (\ref{equality}) uniquely fixes the outer radius of the
CDAF, $\ro$.  It also implies an observationally interesting
correlation between the luminosity and X-ray spectrum of a source.

The simplest possibility we can consider is that $\eta_c=1$, which
corresponds to all the convected energy being radiated by the CDAF.
This may apply to those objects in which the CDAF is formed by the
``evaporation'' of a thin accretion disk, as has been discussed in the
context of ADAFs by several authors (e.g., Meyer \& Meyer-Hofmeister
1994; Narayan, McClintock, \& Yi 1996; Honma 1996; Meyer, Liu, \&
Meyer-Hofmeister 2000).  If a CDAF is surrounded by a thin disk, and
if $\eta_c<1$, the convective luminosity impinging on the inner edge
of the thin disk would be very much greater than that needed to unbind
the disk material.  One could imagine that $L_c$ would then ``eat
away'' the inner edge of the disk until the condition $L_{\rm CDAF}
\approx L_c$ is roughly satisfied.  This condition is not guaranteed,
however, since another possibility is that $\eta_c<1$ and the excess
convective energy is used to eject a substantial fraction of the mass
evaporated from the disk.

When there is no outer disk, it is unlikely that $\eta_c$ will be
close to unity.  For example, if a black hole accretes spherically
from the ISM of a galaxy, as is believed to occur in, e.g., Sgr A* at
the center of our Galaxy (e.g., Melia 1992; Narayan, Yi, \& Mahadevan
1995) or elliptical galaxies at the centers of cooling flows (Fabian
\& Rees 1995; Mahadevan 1997; DM; Quataert \& Narayan 1999b), both
$\ro$ and $\rho_{\rm out}$ are determined by the properties of the
host galaxy.  It would then require considerable fine-tuning for
$L_{CDAF}$ to be equal to $L_c$.  It is more likely that we will have
$L_{\rm CDAF} < L_c$, i.e $\eta_c<1$.  The unradiated energy
$(L_c-L_{CDAF})$ will be transported out, heating the external medium
and perhaps driving some of the gas away in an outflow.

The bremsstrahlung luminosity of a CDAF is given by
\begin{equation}
L_{\rm CDAF} = 4\pi R^{3}_{\rm S} \int_1^{r_{\rm out}} A\rho ^{2}
T^{1/2} r^{2}dr, \qquad A\approx5\times10^{20}~{\rm ergs \ cm^3 \ g^{-2} \ s^{-1} \ K^{-1/2}}.
\end{equation}
Performing the integral and equating $L_{\rm CDAF}$ to the convective
luminosity, we may solve for $\rho _{0}$:
\begin{equation}
\rho _{0} \approx \frac{\eta_c \epsilon _{\rm c}c^{3}}{2 T_0^{1/2} R_{\rm S}A}
{r^{-3/2}_{\rm out}} \approx {10^{-3}\over mr_{\rm out}^{3/2}}
\left({\eta_c\epsilon_{\rm c}\over10^{-2}}\right) ~{\rm g\,cm^{-3}}.
\end{equation}
\noindent
Substituting this back into the expression for $L_{\rm CDAF}$ we find
that the bremsstrahlung luminosity of the CDAF satisfies

\begin{equation}
l_{\rm CDAF}\equiv \frac{L_{\rm CDAF}}{L_{\rm Edd}} \approx 
\ro^{-3/2} \left(\eta_c\epsilon_c \over 10^{-2}\right)^2, \label{lum}
\end{equation} where $L_{\rm Edd}=1.25\times10^{38}m ~{\rm
erg\,s^{-1}}$.  For $\ro \gg 1$, we see that the luminosity of a CDAF
is $\ll L_{\rm Edd}$.  Note that equation (\ref{lum}) can also be
expressed as a relationship between the accretion rate and outer
radius of the CDAF: 
\begin{equation} 
\dot m \equiv {\dot M \over \dot
M_{\rm Edd}} \approx 10 \ro^{-3/2} \left(\eta_c\epsilon_c \over
10^{-2}\right), \label{mdot} 
\end{equation} 
where $\dot M_{\rm Edd} = L_{\rm Edd}/0.1 c^2 \approx 10^{18} m$ g
s$^{-1}$ is the Eddington accretion rate.

A straightforward calculation (ignoring the frequency dependence of
the Gaunt factor) shows that the bremsstrahlung spectrum from a CDAF
consists of three parts:
\begin{eqnarray}
\nu L_{\nu} & \propto & \nu, \qquad h\nu \ll kT_{\rm min}, \\ & \propto
& \nu^{-3/2}, \qquad kT_{\rm min} \ll hv \ll kT_{\rm max}, \\ &
\propto & \exp(-h\nu/kT_{\rm max}), \qquad kT_{\rm max} \ll h\nu,
\end{eqnarray}
where $T_{\rm min} = T(\ro)$ is the minimum electron temperature in
the flow, obtained at the outer radius, and $T_{\rm max}$ is the
maximal electron temperature, obtained close to the black hole.  If we
include the Gaunt factor, the slope is shallower than unity for
$h\nu\ll kT_{\rm min}$.

The peak of the bremsstrahlung spectrum occurs at
\begin{equation}
h\nu_{\rm peak} \sim kT_{\rm min}=\frac{kT_{0}}{r_{\rm out}}.
\end{equation}
\noindent
Thus the position of the spectral peak depends on the outer radius.
Since the luminosity of the CDAF is also related to the outer radius
(eq. 5), we have the following relation between the luminosity and the
location of the peak in the bremsstrahlung spectrum:
\begin{equation}
h\nu_{\rm peak} \approx 10 \left({l_{\rm CDAF}\over10^{-6}}\right)
^{2/3} \left(\eta_c\epsilon_c \over 10^{-2}\right)^{-4/3}\, {\rm
keV}. \label{corr}
\end{equation}

\section{Detailed Spectra}
\label{sec:Detailed Spectra}

\noindent
The detailed spectra presented in this section were computed using a
global dynamical model which includes a sonic transition close to the
black hole (Quataert, in preparation).  The model corresponds to a
viscosity parameter $\alpha=0.03$ and has $\epsilon_c = 0.0045$.  The
profiles of density and temperature in this model are quite close to
the self-similar form assumed in the previous section.

In addition to bremsstrahlung emission, the spectral calculations
presented here include synchrotron radiation and Compton scattering,
assuming thermal electrons.  These processes have been included by the
methods described in Narayan, Barret \& McClintock (1997).  Cooling
from atomic processes has also been included, which can be important
for $T<10^8$ K (i.e $r>10^4$).

In each model, the electron temperature $T_{\rm e}$ was computed
self-consistently at each radius.  Specifically, we solved an energy
equation as a function of $r$ and ensured that the heating, cooling
and energy advection of the electrons are in balance. For the heating,
we assumed that a fraction $\delta$ of the viscous dissipation goes
directly into the electrons.

\subsection{Models with a Pure CDAF}
\label{subsec:Pure CDAF}

Figure 1 shows model spectra corresponding to a CDAF around a $10^8
M_{\odot}$ black hole.  All the models shown have $\eta_c=1$ and
either $\delta = 0.01$ or $\delta = 0.5$.  The various curves
correspond to different values of the outer radius, from $\ro=10^6$
down to $300$; the corresponding values of $\dot m$ are given in the
caption to the figure.  (The precise values of $\dot m$ depend on
details such as the values of $\alpha$ and $\epsilon_c$, and the
global model of the gas dynamics, and probably should not be taken too
literally.  The trends are, however, likely to be robust.)

The main features discussed in \S2 are clearly seen in Fig. 1.
Consider first the models with $\delta=0.01$, shown by dashed lines.
We see that the X-ray spectrum is dominated by bremsstrahlung emission
and consists of two distinct power-law segments on either side of a
peak.  The location of the peak is highly correlated with the
luminosity, and follows the scaling given in equation (10).

At low frequencies, in the radio/submm band, there is a weak secondary
peak in the spectrum which arises from synchrotron emission by the hot
electrons.  (This emission was computed assuming that the magnetic
pressure in the plasma is a tenth that of the gas, cf. Quataert \&
Narayan 1999c.)  The synchrotron peak in the models is much weaker
than in corresponding ADAF models (Narayan, Mahadevan \& Quataert
1997), where the synchrotron emission is often stronger than the
bremsstrahlung emission.  This is because the synchrotron emission
comes from relativistic electrons close to the black hole, whereas the
bremsstrahlung emission is from large radii.  For a given outer
density (and hence bremsstrahlung emission), the electron density at
small radii in a CDAF is much smaller than that in an ADAF, causing
the synchrotron emission in the CDAF to be strongly suppressed.  This
point was made by DM and QN in the context of winds from ADAFs, which
also cause a reduction in the central density.

For higher values of $\dot{M}$ (or luminosity), Compton-scattering of
synchrotron photons becomes important; this fills in the valley
between the synchrotron and bremsstrahlung peaks.

The ions in the above models are close to virial: $T_i\sim10^{12}$~K
near the black hole.  The electrons are comparatively cool, with a
maximum temperature $T_{e,max}<10^{10}$~K.  Models with $\delta=0.5$
(solid lines in Fig. 1) have much stronger electron heating and
therefore significantly hotter electrons: $T_{e,max}\sim10^{11}$
K.  This causes the synchrotron peak to become much more pronounced
compared to the $\delta=0.01$ models, though the emission is still
less than in an equivalent ADAF model.  The increased synchrotron
emission and the hotter electrons both lead to an enhanced
contribution from Compton scattering.  Nevertheless, the peak of the
spectrum is still dominated by bremsstrahlung emission, and so the
scaling of the position of the peak with luminosity survives.


Figure 2 shows spectra for $\delta = 0.01$ and $\delta = 0.5$ for
several values of the convective efficiency, $\eta_c$: the two sets of
models correspond to $\dot m = 10^{-4}$ (solid lines) and $10^{-6}$
(dashed lines).  In all the model sequences the synchrotron emission
in the radio is relatively unaffected by changes in $\eta_c$; this is
because the synchrotron radiation arises primarily at small radii and
depends only on $\dot m$ and $\delta$. For the large $\dot m$ model,
varying $\eta_c$ also has little effect on the X-ray spectrum.  This
is because most of the X-ray emission arises from small radii via
inverse Compton scattering, rather than from large radii via
bremsstrahlung.  By contrast, for smaller $\dot m$ (and/or smaller
$\delta$), inverse Compton emission is less important.  The X-ray
emission is then dominated by bremsstrahlung, which decreases with
decreasing $\eta_c$.

Figure 3 shows the predicted X-ray photon index $\Gamma$ as a function
of X-ray luminosity in Eddington units for $\delta=0.5$.  Four sets of
models are shown: $\eta_c=1, ~0.5, ~0.2, ~0.1$.  The 2--5 keV spectral
indices were calculated by comparing the model luminosities at the two
ends of the range, 2 keV and 5 keV, respectively.  Note the large
increase of $\Gamma$ with decreasing luminosity.  This arises because
the peak in the spectrum moves to lower energies with decreasing
luminosity (cf. eq. 10).  Because the value of $\eta_c$ is unknown,
there is some uncertainty regarding the value of the luminosity at
which the transition in the value of $\Gamma$ occurs .  For the range
of $\eta_c$ considered here, covering an order of magnitude, we find
that for $L_X/L_{Edd}\gtrsim10^{-7}$ the X-ray spectrum is hard, with
a photon index $\Gamma\sim2$, while for luminosities
$L_X/L_{Edd}\lesssim10^{-9}$ the spectrum is significantly softer,
with $\Gamma\sim3.5$.

\subsection{CDAF vs. ADAF}

Figure 4 shows a comparison of CDAF and ADAF spectral models for
$\delta = 0.01$ and $\delta = 0.5$.  The accretion rates in the models
are adjusted so that the $1$ keV X-ray luminosities are equal to
either $10^{37}$ or $10^{40}$ erg s$^{-1}$.  In order to uniquely fix
the models, we set $\eta_c = 1$ in the CDAF models and choose the
outer radii of the ADAF modles so that local viscous dissipation
balances radiative cooling at this radius ($f \lesssim 1/2$ in the
notation of Narayan \& Yi 1994).

For small $\delta$, Figure 4 shows that the CDAF and ADAF spectra are
quite different.  The X-ray emission in the CDAF is dominated by
bremsstrahlung, while the emission in the ADAF is dominated by inverse
Compton emission.  In addition, the CDAF has a much lower synchrotron
flux in the radio, because the density of gas close to the black hole
is much smaller.

For very low luminosities ($< 10^{-8} L_{Edd}$), the difference
between the CDAF and ADAF models at large $\delta$ is similar to that
at small $\delta$.  By contrast, for luminosities $> 10^{-6} L_{Edd}$
the CDAF and ADAF spectra are very similar if $\delta$ is large.  This
is because the entire spectrum is produced over a small range of radii
close to the black hole.  Fixing the $1$ keV X-ray luminosity
determines the properties of the flow in this region and thus the
entire spectrum is rather similar.

\section{Discussion}
\label{sec:Discussion}

The main results of this paper are given in equations (6)--(10) and in
the Figures.  Equation (10) shows that a CDAF has a peak in $\nu
L_\nu$ in the X-ray band, whose position is correlated with the
Eddington-scaled luminosity.  Correspondingly, the spectral index of
the X-ray spectrum varies with luminosity.  At high luminosities, the
peak in the spectrum is at higher energies than the typical band in
which observations are carried out (few keV), and the spectrum is hard
with a photon index $\Gamma$ of order 2 (Fig. 3).  At lower
luminosities, the peak shifts to lower energies, and the observed band
would correspond to energies above the peak.  The spectrum then
becomes very soft, with $\Gamma$ taking values of order 3.5.

There is some uncertainty in the quantitative details predicted here
because the values of two parameters are not well determined.  The
parameter $\epsilon_c$ measures the fraction of the rest mass energy
of the accreting gas that is carried outward by convection:
$L_c=\epsilon_c\dot Mc^2$.  Our calculations correspond to
$\epsilon_c=0.0045$, as determined from a global CDAF model with
viscosity parameter $\alpha=0.03$ (Quataert, in preparation).  Other
values of $\alpha$ would give other estimates of $\epsilon_c$, though
the uncertainty in $\epsilon_c$ is probably no more than a factor of a
few either way.

A more serious uncertainty is in the parameter $\eta_c$, which
measures the fraction of the convected luminosity $L_c$ that is
radiated by the CDAF at large radii as thermal bremsstrahlung
emission: $L_{CDAF}=\eta_cL_c=\eta_c\epsilon_c\dot Mc^2$.  Figure 2
shows results spanning a range of $\eta_c$ from 0.01 to 1.  If
$\eta_c$ is as small as 0.01 or even smaller, an accreting black hole
would have a very small X-ray luminosity from thermal bremsstrahlung
emission and would be extremely difficult to observe.  This might
explain highly underluminous galactic nuclei such as that in our own
Galaxy.  For models with low $\eta_c$, a
relevant question is what happens to the part of the energy convected
outwards that is not radiated, ($[1-\eta_c] L_c \approx L_c$).  If
this energy is transported out to an external medium and radiated
there, that contribution may dominate the observed spectrum (assuming
the beam of the telescope is larger than the size of the emitting
region).  The luminosity of the source would then be larger than we
have predicted and the spectrum would be softer.

The spectral models of CDAFs presented here are qualitatively similar
to the ADIOS (ADAF + wind) models presented by DM and QN.  The
relative prominence of bremsstrahlung emission in the X-ray band and
the suppression of synchrotron emission in the radio (as suggested by
observations; see DM) are characteristic of density profiles flatter
than the ADAF scaling of $\rho \propto r^{-3/2}$ (be they CDAF or
ADIOS).  In fact, the qualitative similarity of ADIOS and CDAF spectra
implies that direct signatures of outflowing gas (and measrements of
outflow mass rates) would be needed to confirm the ADIOS model.  The
primary difference between the CDAF and ADIOS models is that we have a
unique radial density profile ($\rho \propto r^{-1/2}$) in a CDAF,
instead of a family of models ($\rho \propto r^{-3/2 + p}$ with
$0<p<1$) in an ADIOS; also, the energy which powers bremsstrahlung
emission at large radii is provided by convective transport from small
radii in a CDAF, rather than by local viscous dissipation in an ADIOS.


All the models presented in this paper correspond to a two-temperature
plasma.  For the models with $\delta=0.5$, however, the temperatures
of the ions and electrons are similar, differing by less than a factor
of 10.  We have computed models with a one-temperature plasma,
assuming that some plasma process other than Coulomb collisions
rapidly equilibrates the ion and electron temperatures (cf Begelman \&
Chiueh 1988).  The results are not very different from those obtained
with two-temperature models with $\delta=0.5$.  Comptonization does,
however, become more important relative to bremsstrahlung, especially
for low values of $\eta_c$.  In addition, the synchrotron emission in
the radio is somewhat more prominent.


An important assumption of our analysis is that the electrons are
thermal.  Because of the low density in a CDAF, thermalization through
Coulomb collisions and synchrotron self-absorption is virtually
non-existent (cf. Mahadevan \& Quataert 1997); it is therefore
possible to retain a power-law distribution of electrons.  Moreover,
such nonthermal acceleration is expected in the collisionless
magnetized plasmas present in CDAFs.  A power law tail of electrons
would significantly modify the predicted synchrotron spectrum (e.g.,
Mahadevan 1998; \"Ozel, Psaltis, \& Narayan 2000).  In addition, the
power-law electrons could contribute to X-ray emission through either
synchrotron or nonthermal inverse Compton emission.  Depending on the
uncertain efficiency of electron acceleration, this nonthermal
emission could be more important than bremsstrahlung emission,
particularly for small values of $\eta_c$.

Models in which thin accretion disks ``evaporate'' to form hot inner
flows have received considerable attention in the context of dwarf
novae (Meyer \& Meyer-Hofmeister 1994), soft X-ray transients in
quiescence (Narayan, McClintock, \& Yi 1996; Honma 1996; Meyer, Liu,
\& Meyer-Hofmeister 2000, Rozanska \& Czerny 2000), and low luminosity
AGN (Lasota et al. 1996; Quataert et al. 1999).  The energy for
``evaporation'' is ultimately thought to originate in the hot inner
flow.  Convection in CDAFs transports a significant amount of energy
to large radii.  CDAFs should thus be much more efficient than ADAFs
at evaporating the outer disk.  The implications of convective energy
transport for theoretical models of the transition from thin disks to
hot flows deserve further investigation.

Acknowledgment: The authors are grateful to an anonymous referee for a
number of suggestions which helped improve the presentation.  This
work was supported in part by grants PHY 9507695 and AST 9820686 from
the National Science Foundation.  GB was supported by a Frank Knox
Memorial Fellowship from Harvard University. EQ is supported by NASA
through Chandra Fellowship PF9-10008, awarded by the Chandra X--ray
Center, which is operated by the Smithsonian Astrophysical Observatory
for NASA under contract NAS 8-39073

\newpage

\begin{figure}
\plotone{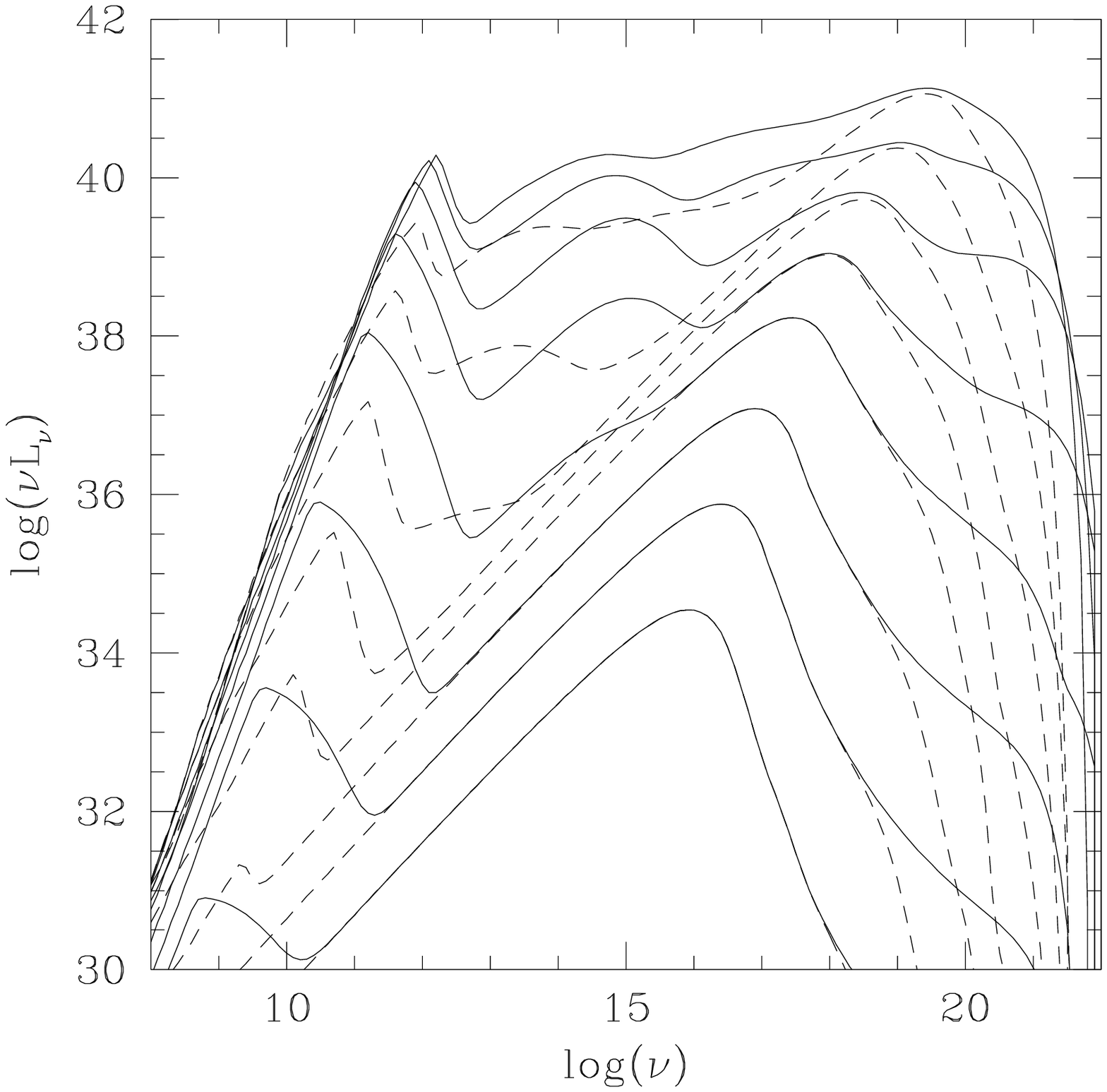}
\caption{Model CDAF spectra for a $10^8 M_\odot$ black hole, taking
$\eta_c = 1$.  Dashed lines are for a fraction $\delta = 0.01$ of the
viscous energy heating the electrons.  From top to bottom, the curves
correspond to models with $(\log\ro,\log\dot m) = (2.5,-3.29)$,
$(3.0,-3.86)$, $(3.5,-4.49)$, $(4.0,-5.19)$, $(4.5,-5.99)$, $(5.0,-7.15)$,
$(5.5,-8.36)$, $(6.0,-9.70)$, respectively.  Solid lines are for $\delta
= 0.5$.  From top to bottom, the curves correspond to models with
$(\log\ro,\log\dot m) = (2.5,-3.38)$, $(3.0,-3.91)$, $(3.5,-4.50)$,
$(4.0,-5.19)$, $(4.5,-5.99)$, $(5.0,-7.15)$, $(5.5,-8.36)$, $(6.0,-9.70)$,
respectively.}
\end{figure}

\begin{figure}
\plottwo{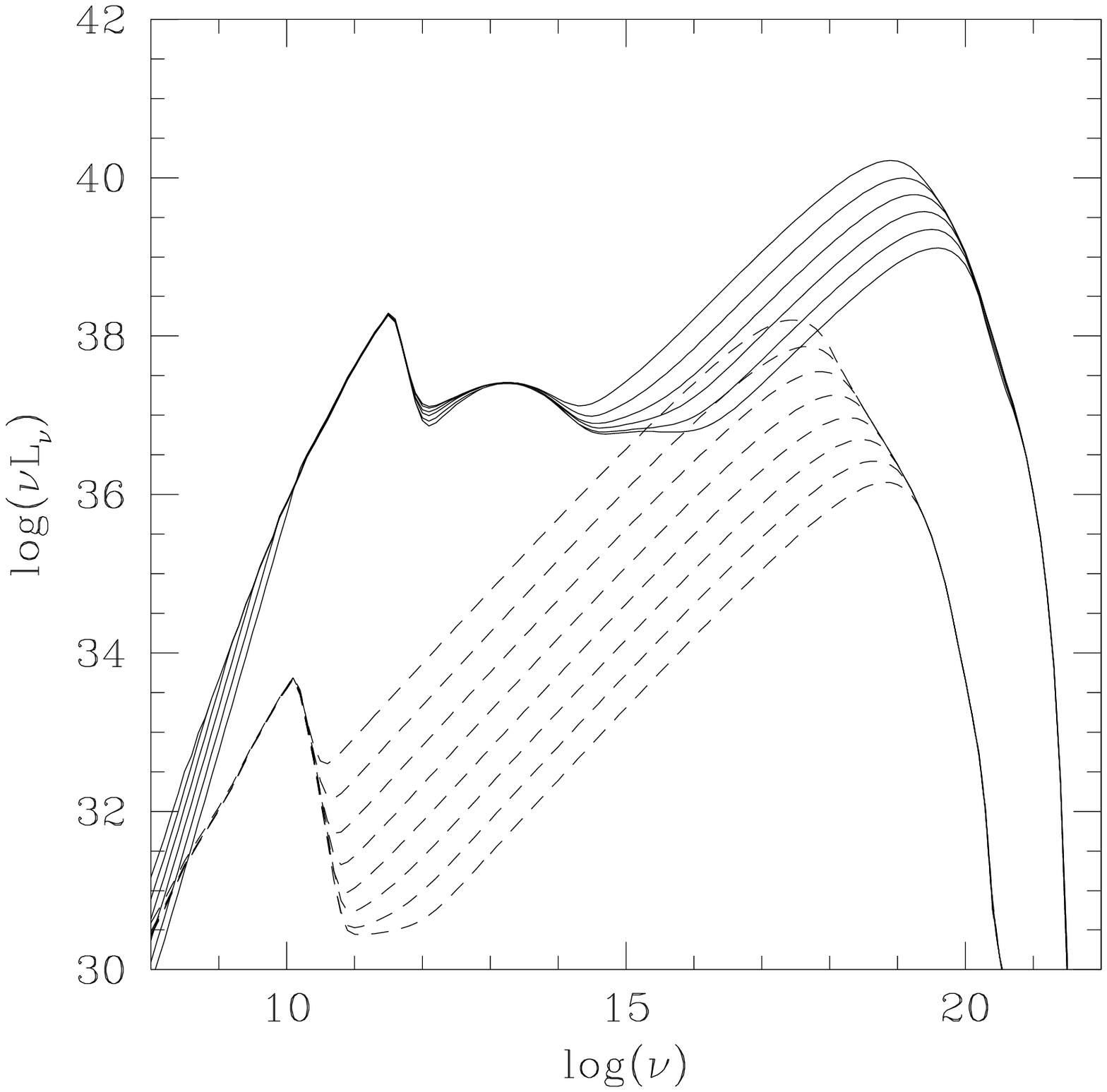}{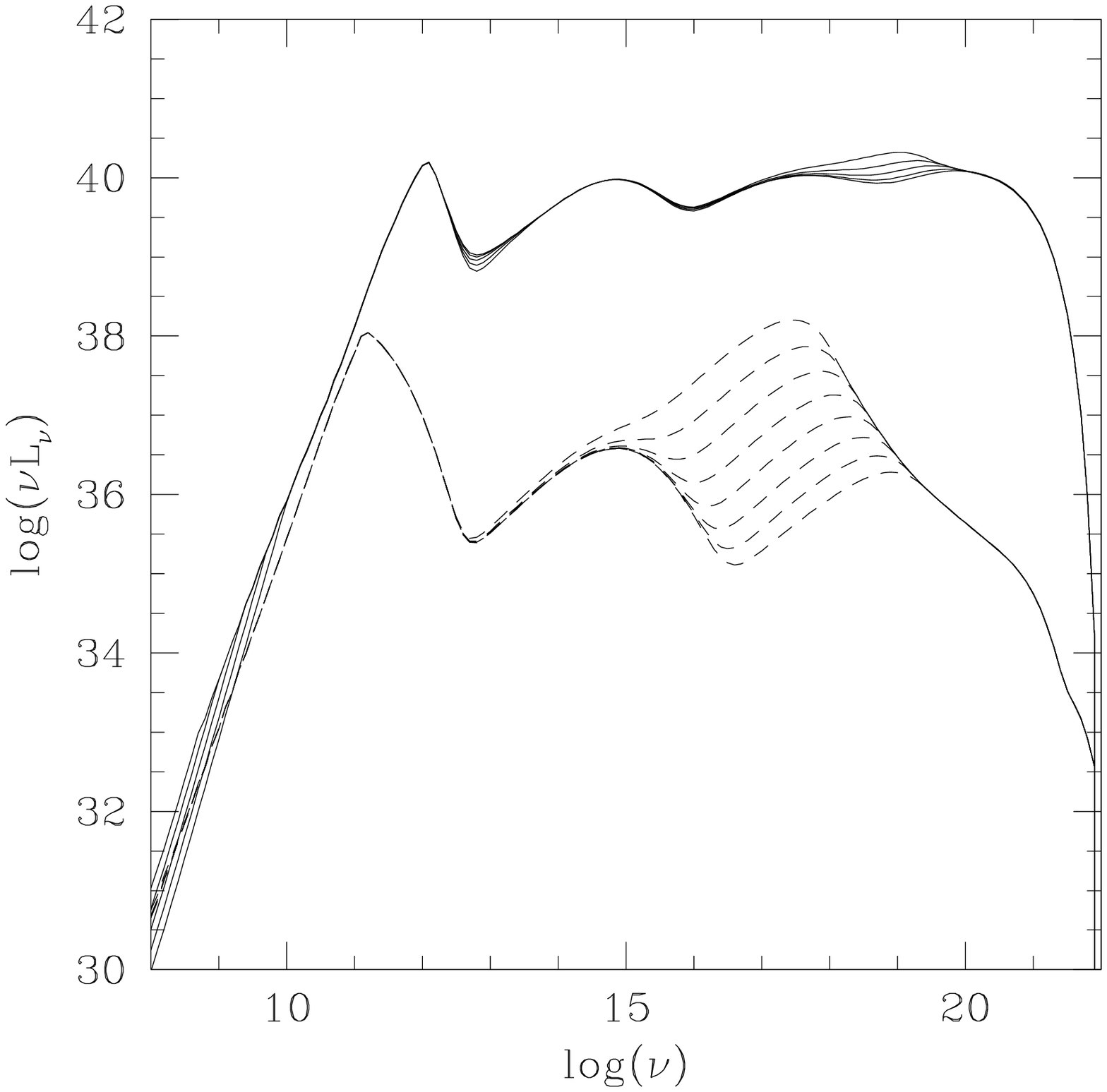}
\caption{Model CDAF spectra for a $10^8 M_\odot$ black hole.  {\it
Left panel:} Models with $\delta = 0.01$.  The solid curves correspond
to $\dot m=10^{-4}$ and, from top to bottom,
$(\log\ro,\eta_c)=(3.1,0.96)$, $(2.9,0.55)$, $(2.7, 0.31)$, $(2.5,0.16)$,
$(2.3, 0.074)$, $(2.1,0.019)$, respectively.  Dashed curves correspond
to $\dot m=10^{-6}$ and, from top to bottom,
$(\log\ro,\eta_c)=(4.5,0.97)$, $(4.3,0.45)$, $(4.1, 0.22)$, $(3.9,0.11)$,
$(3.7, 0.056)$, $(3.5,0.029)$, $(3.3,0.015)$, $(3.1,0.0080)$,
respectively.  {\it Right panel:} Models with $\delta = 0.5$.  The
solid curves correspond to $\dot m=10^{-4}$ and, from top to bottom,
$(\log\ro,\eta_c)=(3.0,0.86)$, $(2.8,0.54)$, $(2.6, 0.34)$, $(2.4,0.20)$,
$(2.2, 0.094)$, respectively.  Dashed curves correspond to
$\dot m=10^{-6}$ and, from top to bottom, $(\log\ro,\eta_c)=(4.5,0.97)$,
$(4.3,0.45)$, $(4.1, 0.22)$, $(3.9,0.11)$, $(3.7, 0.059)$, $(3.5,0.032)$,
$(3.3,0.018)$, $(3.1,0.011)$, respectively.}
\end{figure}

\begin{figure}
\plotone{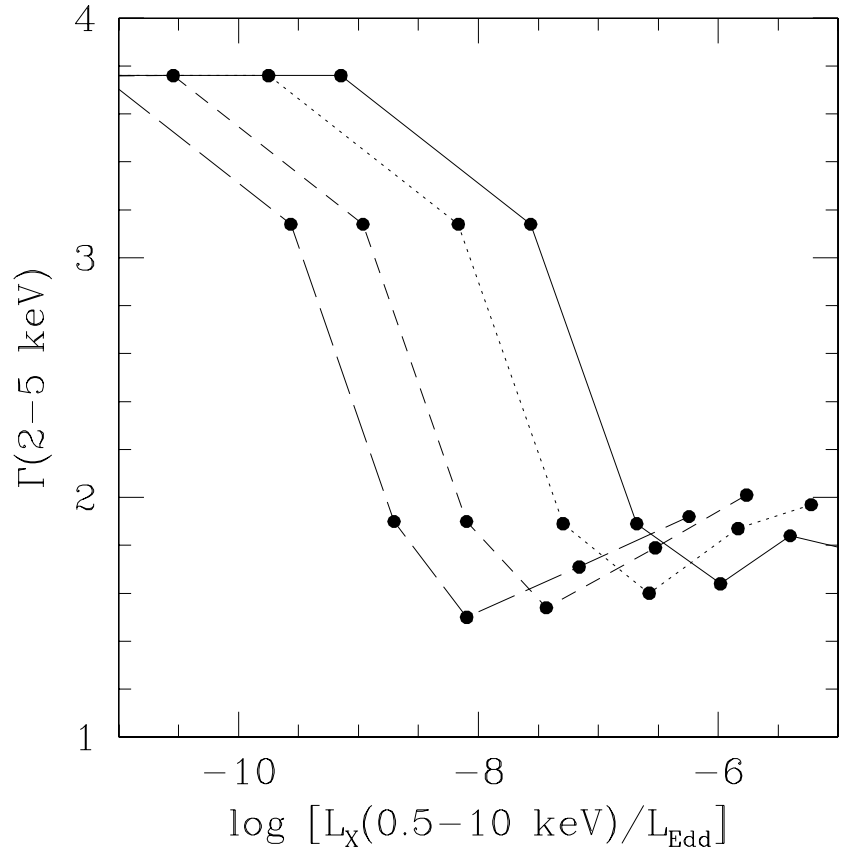}
\caption{X-ray photon index vs. X-ray luminosity for models with
$\delta = 0.5$.  Solid, dotted, dashed, and long-dashed curves
correspond to $\eta_c = 1, 0.5, 0.2$, and $0.1$, respectively.}
\end{figure}




\begin{figure}
\plottwo{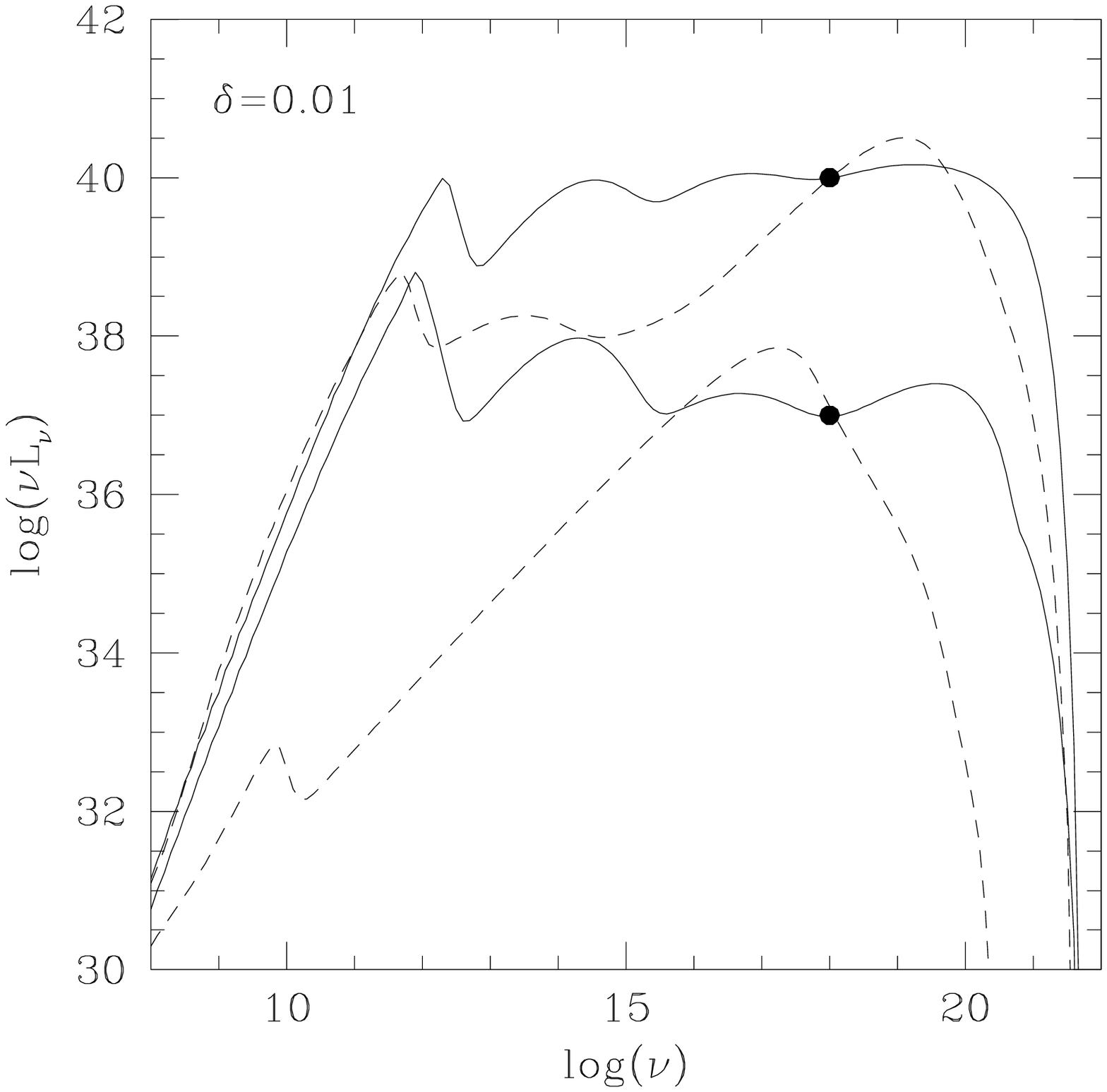}{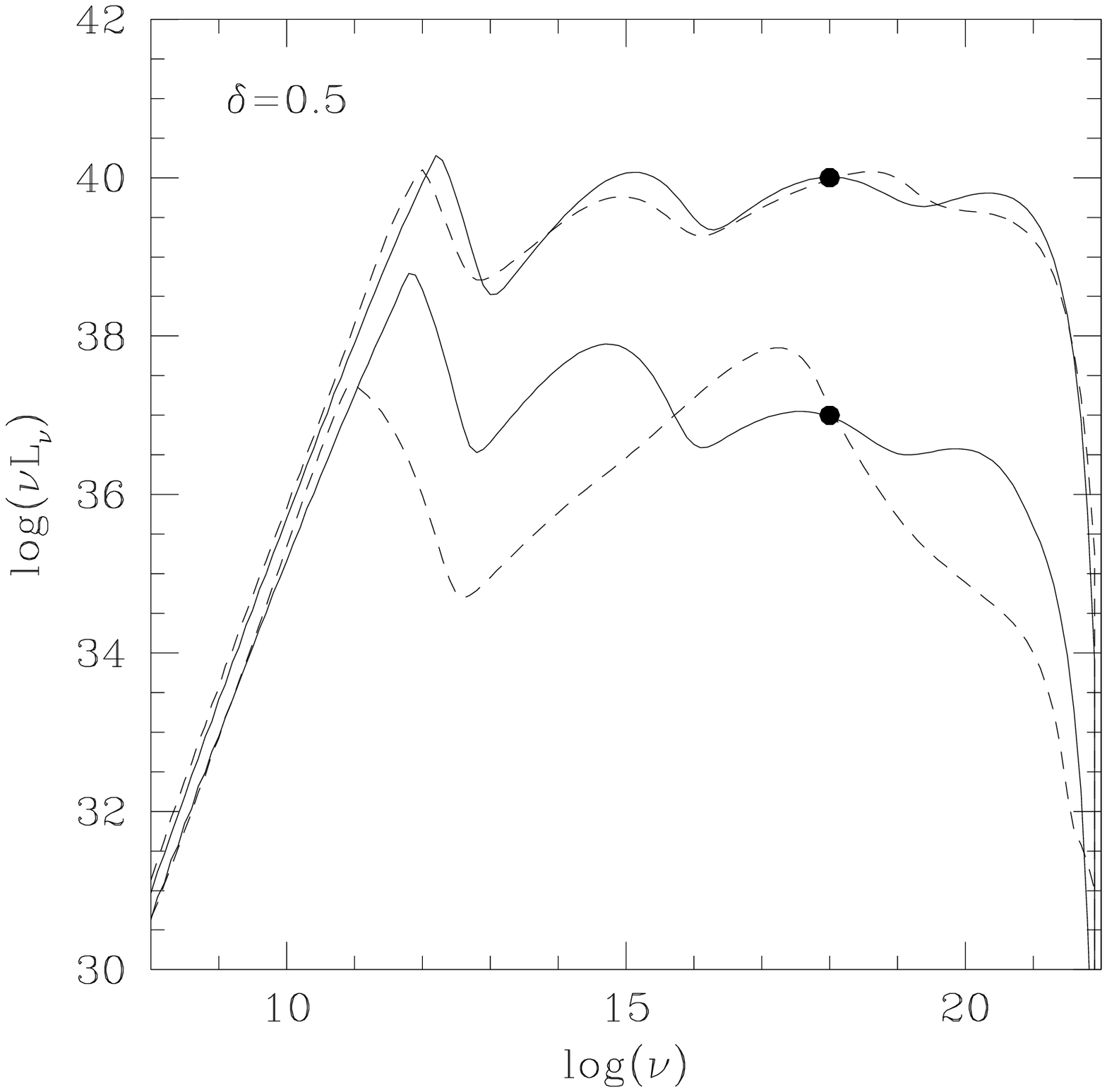}
\caption{Comparison of ADAF and CDAF model spectra; $\eta_c = 1$ for
the CDAF models.  The accretion rates in the models have been adjusted
so that the 1 keV luminosities are equal to either $10^{37} ~{\rm
erg\,s^{-1}}$ or $10^{40} ~{\rm erg\,s^{-1}}$ (solid circles).  {\it
Left panel:} Models with $\delta = 0.01$.  The ADAF models (solid
curves) correspond, from above, to $(\log\ro,\log\dot m)=(4.9,-2.74)$
and $(5.2,-3.64)$, respectively, and the CDAF models (dashed curves)
correspond to $(\log\ro,\log\dot m)=(2.9,3.75)$ and $(4.7,-6.37)$,
respectively.  {\it Right panel:} Models with $\delta = 0.5$.  The
ADAF models (solid curves) correspond to $(\log\ro,\log\dot
m)=(5.5,-3.53)$ and $(5.6,-4.20)$, respectively, and the CDAF models
(dashed curves) correspond to $(\log\ro,\log\dot m)=(3.3,-4.26)$ and
$(4.7,-6.37)$, respectively.}
\end{figure}

\end{document}